\documentclass[fleqn,twoside]{article}

\usepackage{espcrc2,epsfig}

\title{Deep Radio Continuum Studies with the SKA: Evolution of Radio
AGN Populations}

\author{C. A. Jackson\address[atnf]{CSIRO Australia Telescope National
Facility, Sydney, Australia \\ Carole.Jackson@csiro.au}}

\begin{document}

\begin{abstract}

Radio emission is insensitive to dust obscuration, and the breadth of
the radio luminosity function ensures that sources are detected over a
wide range of redshifts at all radio flux densities. As a result,
radio continuum observations are an efficient and unbiased probe of
both nuclear (AGN) and star-forming activity over all cosmic epochs.

The SKA's ``ultra-deep radio continuum surveys'' will provide the
answers to at least three key astrophysical questions: (i) the
star-formation history of the Universe, (ii) the evolution of the
low-power end of the radio galaxy luminosity function and (iii) the
relationship between the radio-loud AGN, star-formation and
radio-quiet AGN phenomena.

In this paper we discuss the AGN science that will be enabled by the
deep radio continuum studies using the SKA. It is important to
recognise that the sub-$\mu$Jy SKA sky will be dominated by
populations other than `radio-loud' AGN.  In this way the SKA will
not only provide an unbiased tracer of the star-formation history of
the Universe but also be able to study the populations of sources we
currently describe as `radio-quiet'.  To illustrate this point we
present simulations of the extragalactic radio sky based from models
of the evolution of the radio luminosity function. From these
simulations we predict typical source distributions and estimate the
natural confusion limit to ultra-faint flux density limits relevant to
the science and design goals of the SKA.

\vspace{1pc}
\end{abstract}

\maketitle


\section{INTRODUCTION}

As a consequence of the broad radio luminosity function (RLF) which
spans many decades in radio power, radio continuum observations at all
flux density limits detect extragalactic sources across a wide
redshift range. Current large-area radio surveys, e.g. the NRAO VLA
Sky Survey (NVSS, \cite{co2}), Faint Images of the Radio Sky at twenty
cm (FIRST, \cite{wb1}) and the Sydney University Molonglo Sky Survey
(SUMSS, \cite{bl1}), cover thousands of square degrees of sky to
mJy-flux density limits. These surveys are the deepest practicable
over large areas of sky given the large amounts of observing time
involved. Surveys to increasingly fainter flux density limits cover
far smaller areas, e.g. the Australia Telescope ESO Slice Project
(ATESP, \cite{pg2}) covers 26 square degrees at 1.4~GHz to $S_{1.4 \rm
\thinspace GHz}$ = 0.4~mJy (5$\sigma$ rms) whereas the VLA
observations of the Hubble Deep Field (north) area covers just 0.35
square degrees \cite{rr1} to $S_{1.4 \rm \thinspace GHz}$ = 0.04~mJy
(5$\sigma$ rms).

The SKA, with its significantly larger collecting areas, will be
capable of large-area sub-$\mu$Jy surveys as well as deep nano-Jy
small-area surveys. The challenge is to determine what the sky might
look like at these faint limits. Firstly it is obvious that the SKA
will be able to compile a complete census of AGN throughout the
Universe, mapping their evolution and reveal the relationship between
the various populations. In turn this will allow us to unravel the
underlying physics of the AGN lifecycle (Jarvis \& Rawlings, Falcke et al, this volume).  The key AGN questions are

- When do the first AGN form? What is their role in the epoch of reionization?

- What drives the evolution of AGN? Is it in-fall of material via
mergers, internally-changing accretion rates?

- What is the typical lifecycle of an AGN? Is it the same galaxies
re-activating?

- How many populations of AGN are there? 

- Is there a continuum of AGN activity, from the super-Eddington
accretors to normal galaxies (and perhaps to even lower-power/low mass
objects)? Are there objects with BH's of a few hundred solar masses?

- What is the interplay (if any) between AGN activity and star formation?

In this contribution we concentrate on how SKA deep continuum surveys
will revolutionise our study of the AGN phenomenon in light of our
current knowledge.

\section{RADIO-LOUD AGN, CONTINUUM SURVEYS AND THE RLF}

Bright continuum radio surveys reveal sources which are powered by
non-thermal (synchrotron) radio emission from AGN embedded within
individual galaxies span a huge range in luminosities. AGN activity is
detected across a huge radio luminosity range from between 10$^{22}$ W
Hz$^{-1}$ to 10$^{28}$ W Hz$^{-1}$ at 1.4~GHz.

At low radio powers we find classes of objects more usually termed
radio-quiet AGN: These include Seyfert galaxies and QSOs. Because
of the huge range of observed luminosities, it can be argued that a
continuum of nuclear radio activity exists - such that all galaxies
may harbour a central MBH and it is only periodical triggering of
activity which gives rise to radio jets and lobes

Radio sources are classified by types, e.g., normal galaxies,
Seyferts, FRI or FRII radio galaxies or quasars, based on their
luminosity, radio morphology, radio spectral index, variability,
optical counterpart, and IR, optical and X-ray spectral features.

The existence of a continuum of radio activity is clear, with the
various source populations overlapping at all but the highest radio
luminosities. This is clearly seen in recent, accurate determinations
of the local radio luminosity function (LRLF, \cite{cc1,ss1}) for
moderate power radio sources (10$^{22}$ to 10$^{24}$ W Hz$^{-1}$
derived from comprehensive galaxy redshift surveys (UGC, 2dFGRS, SDSS)
matched to large-area radio surveys (NVSS, FIRST, SUMSS).  At lower
radio luminosities, the LRLF can only be compiled by observing
individual objects for long integration times. At the high luminosity
end the space density of objects is small at all epochs, so that all
that can be determined is the sparsely-sampled (evolving) RLF from
large-area samples (e.g. \cite{w04}).  Moreover the
high luminosity end of the LRLF can {\it only} be derived from
evolutionary fits to the source count, as it cannot be directly
determined due to the very low local space density of these objects.

In terms of source types, powerful extragalactic radio sources are a
mixture of radio-loud galaxies, quasars and BL Lac objects -- all
members of the Active Galactic Nuclei (AGN) class of objects. At the
zeroth level, radio galaxies can be classified as either
Fanaroff-Riley class I or II (FRI or FRII): These are distinguished by
the distribution of bright knots of radio emission and collimation
features \cite{fr1}. In general, FRIs are of lower radio power than
FRIIs, although there is considerable overlap in the radio luminosity
function of the two classes. The current paradigm for radio-loud AGN
attributes FRI and FRII radio galaxies as the `parent' populations
of the great majority of the observed powerful AGN (e.g. \cite{up1},
such that quasars, blazars and BL Lac objects are manifestations of
FRI and FRII radio galaxies aligned close to our line of sight, with
their core radio emission Doppler-beamed (\cite{jw1} and references
therein).

Typically, FRI and FRII radio sources have linear dimensions of the
order of between one and a few hundred kilo-parsecs. Their radio
structure shows a pair of oppositely directed radio lobes which are
sometimes joined to the radio core via a pair of collimated, knotty
jets of bright radio emission. The radio emission from the lobes of
FRI and FRII radio galaxies is usually optically thin with a radio
spectral index around $-$0.7, although these sources may also have a
self-absorbed flat-spectrum core. Compact radio quasars, and radio
galaxy cores, typically have angular dimensions much less than an
arc-second, and due to self-absorption have flat radio spectra. The
compact radio sources (quasars and blazars) are often variable across
a wide waveband (radio-to-UV) and their host galaxies may show broad
optical/UV emission line spectra.

The study of the evolution of the most powerful radio-loud AGN is
based on the RLF. It is well established that powerful evolution is
required, for both these objects, and radio-quiet QSOs: The form of
this evolution -- whether luminosity evolution \cite{bo1} or
luminosity-density evolution \cite{dp1} is not yet determined, but
pure luminosity evolution infers source lifetimes which are unphysical
\cite{hr1}. Observations at SKA sensitivities are required to detect
low power radio sources at high redshift; this will determine whether
all radio-loud AGN co-evolved or if we are seeing multiple populations
undergoing bursts of AGN activity.

Classification of individual radio sources is often hampered by the
difficulty of optical/IR/UV follow-up: if the optical host is
identifiable it can still be too faint for 8-m class
spectroscopy. Fortunately cross-waveband detection may not be a
requirement for the SKA as the distinction between AGN and/or
star-formation activity is best made via VLBI-scale imaging of $\mu$Jy
sources.

\subsection{Radio sources and the SKA}

Significant progress has been made in understanding the general radio
AGN phenomenon from both

\begin{description}

\item (i) wide and shallow sky surveys, cataloguing up to 100 sources
per square degree to about 1~mJy (e.g. FIRST), and

\item (ii) small and deep pencil beam surveys ($<<$1 degree areas), to
very faint flux densities – e.g. the Hubble Deep Field North and
other fields \cite{fk1} and the Hubble Deep Field South \cite{no1}.

\end{description}

The primary gain of the SKA will be its ability to reach deep
sensitivity limits and probe the enormous span of AGN radio powers
across a wide redshift range. This will remove the luminosity-redshift
(P-z) degeneracy which plagues flux density limited samples of today:
at significant redshifts (z$>$1) only the highest-power sources are
detected \cite{bu1}.

Multi-wavelength (optical/UV) imaging and spectroscopy of sources in
the wide and shallow surveys have revealed that the vast majority of
radio sources stronger than a few mJy at 1.4~GHz are powerful radio
galaxies or quasars.  Moreover, due to the steepness of the high radio
power RLF, coupled with strong cosmic evolution between z=1 and z=5,
the {\it entire} population of high power radio AGN is sampled above a
few mJy: below this flux density limit the nature of the radio source
population changes dramatically. Small, deep pencil beam surveys
reveal that below about 1~mJy at 1.4~GHz, the differential radio
source count flattens: both the star forming galaxy population and
galaxies with low power AGN contribute. The contribution from, and
physical properties of the star forming galaxies is discussed in
van der Hulst et al. (this volume).

\section{THE SKA EXTRAGALACTIC CONTINUUM SKY}

The SKA will offer significant improvement in sensitivity so that it
will be possible to detect radio emission as weak as 10~nJy within
$\sim$ 1000 hours integration time. In this way, the SKA will probe a
huge part of the very broad RLF at all epochs and the misnomers of
`radio-loud' -- `radio-quiet' may be buried forever.

High resolution, multi-wavelength observations will be critical to
distinguish those extragalactic radio sources with AGN emission
processes.  Whilst nothing is known about the radio source count below
about 30$\mu$Jy, some estimates can be made by extrapolating the RLF
to low luminosities (10$^{18}$ W Hz$^{-1}$) and adopting evolution
models which match the observed evolution of the more powerful
sources.

\subsection{Simulating the radio source sky}

Simulations of the extragalactic radio sky, based on number density
predictions from models of the evolution of the radio luminosity
function allow us to predict typical source distributions and natural
confusion limit to projected SKA flux density limits. We show some
examples here for a lambda-dominated `WMAP' cosmology
($\Omega_{m}$=0.23, $\Omega_{\Lambda}$= 0.73 and $H_{0}$= 71 km
s$^{-1}$ Mpc$^{-1}$).

We start by assuming that the radio sky comprises three populations of
sources, namely FRI, FRII and star-forming galaxies, and the LRLF and
evolution for each can be determined: For the star-forming galaxy
population we adopt the local radio luminosity function from the
2dFGRS-NVSS galaxy sample at 1.4~GHz \cite{ss1}. We adopt
parameterised number-density luminosity evolution for the star-forming
population as determined from the HDF-N and SSA13 fields \cite{hp1}.

The local radio luminosity functions and evolution for the FRI and
FRII radio galaxies are derived using the methodology of Jackson \&
Wall (1999). In summary, we fit exponential luminosity-dependent
density evolution (LDDE) to the observed source count at 151~MHz. We
find a best-fit to the observed 151~MHz source count for LDDE of the
FRII population, $\phi_{z} = F(P,z).\phi_{0}$, where $\phi_{0}$ is the
LRLF. For the function $F(P,z)$ we fit $M_{max}$=15.70, $z_{c}$=7.425,
$P_{1}$=25.94 $P_{2}$=27.94 for the FRII population, coupled with mild
($M = -3.06$) density evolution of the FRI population.

From the three evolving RLFs we can produce simulated sky regions at a
range of SKA frequencies: At 1.4~GHz we use the evolutionary fit at
151~MHz plus `beaming factors' to account for the quasar and BL Lac
populations (Fig.1). For a single sky simulation we randomly position
each source within a specified sky area. Each source is then randomly
oriented, both on and into the plane. FRI and FRII radio galaxies are
randomly assigned a total intrinsic size (lobe-to-lobe) between 50 kpc
and 800 kpc (as observed for the powerful 3CR sources, \cite{lr1}),
assuming no source size evolution with redshift. FRIs and FRIIs are
modeled as double-lobe structures, with each lobe being half the
intrinsic total size adjusted for source orientation and redshift,
i.e. foreshortened and flattened. Doppler-beamed versions of the FRI
and FRIIs are set to be point sources. Star-forming galaxies are
modelled as a circular disks with a randomly assigned intrinsic disk
diameter between 10 and 100 kpc: these sizes are generous given the increasing evidence that spiral disks at high redshift are significantly smaller (e.g.
\cite{gi1}). These disks are oriented randomly on the plane of the sky. Examples are shown here (Fig. 2), a wider
range of simulated sky images, at 151~MHz, 325~MHz and 1.4~GHz, are
available at http://www.atnf.csiro.au/$\sim$cjackson/lowfreq\_sky.

We can predict the number of sources per square degree and the
fraction of sky covered by each population at 1.4~GHz as shown in
Table~\ref{surf1400}. We estimate the minimum resolution which would
give an average inter-source spacing of 10 primary beams on the sky,
allowing for the varying set of source sizes in a single sky
simulation. Between $S_{1.4 \rm \thinspace GHz}$= 10~$\mu$Jy and
100~nJy the minimum resolution requirement to avoid natural confusion
ranges from 0.86 arc-seconds to 0.07 arc-seconds. Also of interest for
deep SKA observations is the natural confusion rate, i.e. the
proportion of sources which overlap each other on the plane of the
sky. We find that this varies from $\sim$10\% of all sources at
10~$\mu$Jy, rising to $\sim$60\% at 100~nJy. This means that at a flux
density limit of $S_{1.4 \rm \thinspace GHz}$= 100~nJy, {\it more than
half} of all detected sources will be in line-of-sight coincidence
with another source.

\begin{figure*}
\begin{center}
\psfig{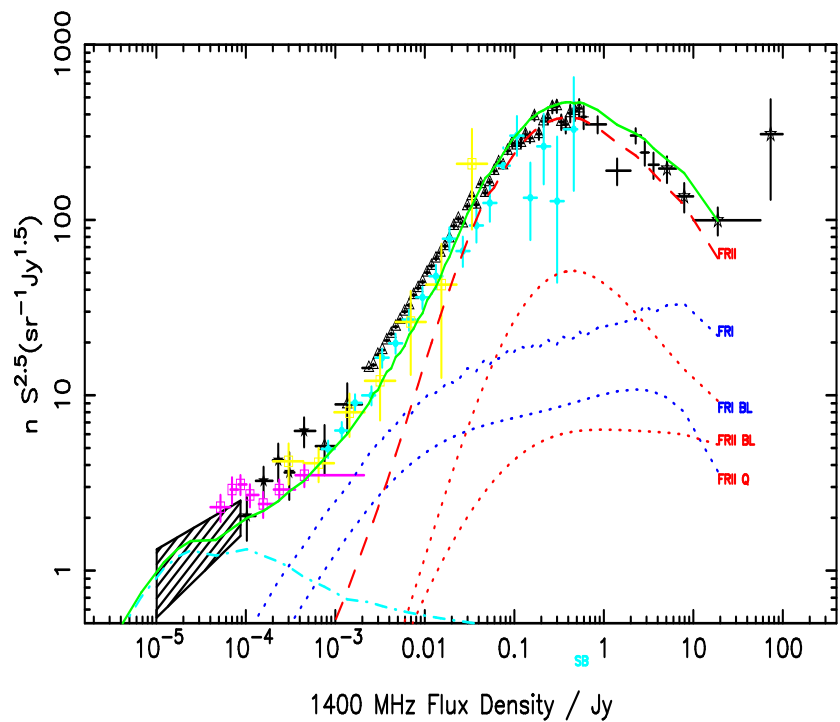}
\caption{Normalised observed and model differential source counts at 1.4~GHz.
The model count comprises 3 populations:
FRII radio galaxies (dashed line), FRI radio galaxies (dotted)
and star-forming galaxies (dot-dash).
Data points are from: $\star$ Bridle et al. (1972), 
$+$ Fomalont et al. (1974),
$\triangle$ White et al. (1997),
$\star$ Mitchell \& Condon (1985), 
$\circ$ Prandoni et al. (2000), 
$\triangle$ Gruppioni et al. (1997), and 
$\ast$ Richards (2000).
The polygon is the count estimate from the P(D) (background-deflection)
analysis of Wall \& Cooke (1975).}
\label{fit1400l}
\end{center}
\end{figure*}

\begin{table*}
\caption{Predicted source surface densities at 1.4~GHz.}
\vspace*{0.2in}
\begin{tabular}{lrrrrrr}
\hline
& \multicolumn{6}{c}{Flux density limit} \\
& \multicolumn{2}{c}{10~$\mu$Jy} &
\multicolumn{2}{c}{1~$\mu$Jy} & 
\multicolumn{2}{c}{100~nJy} \\
&  N / &  Cover & N / &  Cover &
 N /  &  Cover  \\
Population & deg$^{2}$  &
fraction & deg$^{2}$  &  fraction &
deg$^{2}$  &  fraction\\
\hline
\\
FRI         &   1,207 & 4.10$^{-2}$ & 4,028 & 0.136    & 10,162 & 0.340 \\
FRII        &     55 & 3.10$^{-3}$ & 56 & 3.10$^{-3}$  &  56 & 3.10$^{-3}$  \\
star-forming &  7,361 & 2.10$^{-3}$ & 52,798 & 2.10$^{-2}$ &  135,806 & 0.05 \\
\\
\hline
\\
Total      & 8,623 & 5.10$^{-2}$  & 56,822 & 0.162 & 146,024 & 0.394 \\
\hline
Minimum \\
resolution  \\
/ arcsecs & \multicolumn{2}{c}{0''.86} & \multicolumn{2}{c}{0''.17} & \multicolumn{2}{c}{0''.07}  \\
\hline
\\
\% of sources \\
overlapped &
\multicolumn{2}{c}{10\%} & \multicolumn{2}{c}{25\%} & \multicolumn{2}{c}{60\%} 
\\
(line-of-sight) \\
\\  
\hline
\end{tabular}
\label{surf1400}
\end{table*}

Table~\ref{pzt1400} shows the source distributions by redshifts: here
we see that both the FRI and FRII populations extend to very high
redshift (z $\sim$8) at $S_{1.4 \rm \thinspace GHz} <$
10~$\mu$Jy. These sources would provide the foreground sources to HI
line-of-sight experiments (Kanekar \& Briggs, this volume and also \cite{ca1}).

\begin{table*}
\caption{Predicted source distribution per square degree at 1.4~GHz.}
\vspace*{0.2in}
\begin{tabular}{lrrrrrrrrrr}
\hline
& \multicolumn{9}{c}{Flux density limit} \\
& \multicolumn{3}{c}{10~$\mu$Jy} &
\multicolumn{3}{c}{1~$\mu$Jy} & 
\multicolumn{3}{c}{100~nJy} \\
Redshift &  {\small N(FRI)} &  {\small N(FRII)} & {\small N(SF)} &
  {\small N(FRI)} &  {\small N(FRII)} & {\small N(SF)} &
  {\small N(FRI)} &  {\small N(FRII)} & {\small N(SF)} \\ 
\hline
\\
z $<$ 1 &  718 &   2 &   2,556 & 1,636 &  2 &  9,759 & 2,570 &  2 &  14,238 \\
z $<$ 3 & 1,145 &  21 &   6,740 & 3,202 & 21 & 42,865 & 6,551 & 21 & 100,453 \\
z $<$ 5 & 1,190 &  42  &  7,361 & 3,644 & 42 & 52,798 & 7,955 & 42 & 135,806 \\
z $>$ 5 &   17 &  13 &      0 &  384 & 13 &     0 & 2,207 & 13 &      0 \\
\hline
\end{tabular}
\label{pzt1400}
\end{table*}

\begin{figure*}
\begin{center}
\psfig{file=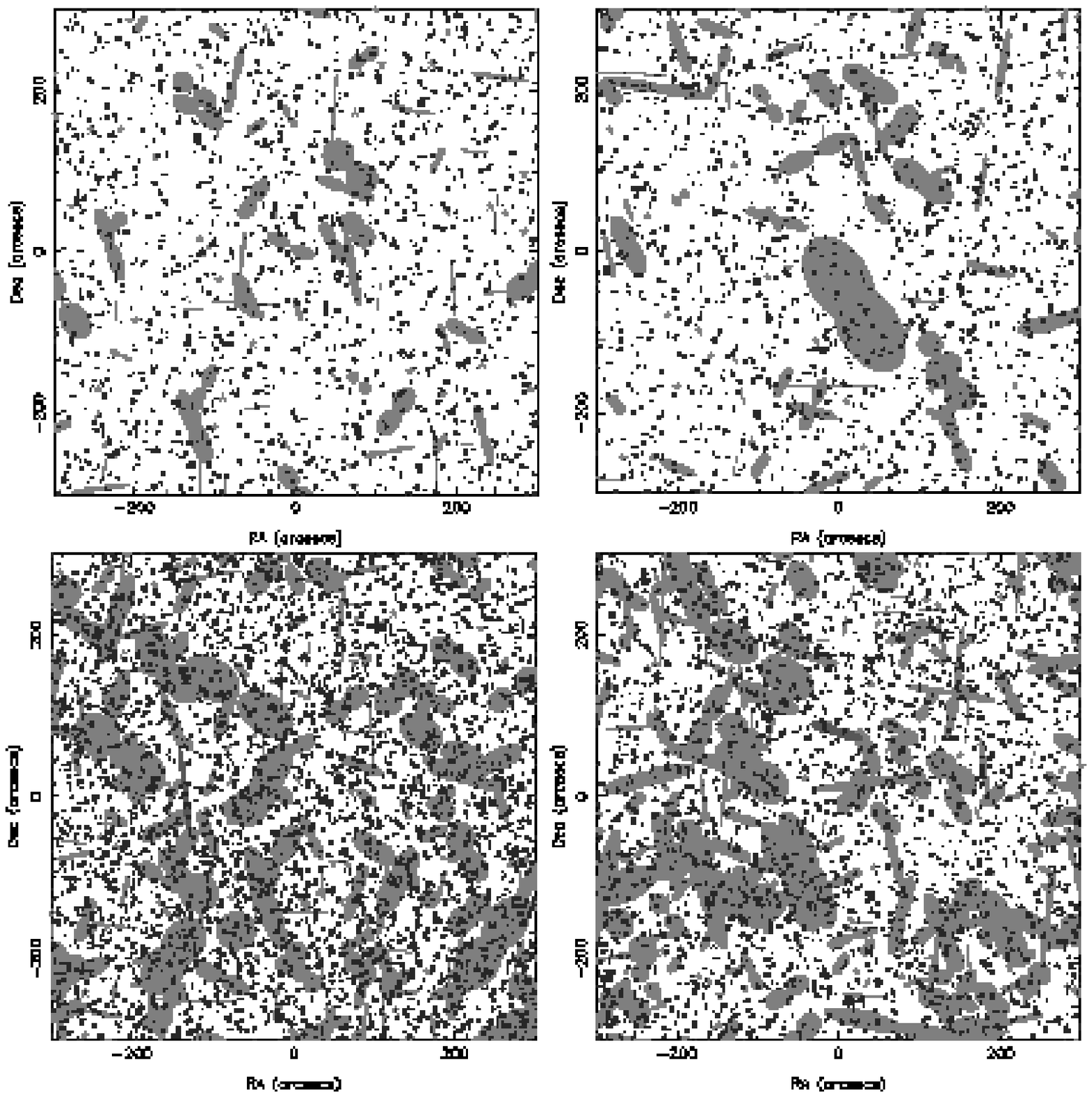,width=13cm,angle=0}
\caption{Simulated sky images at 1.4~GHz, showing a 10 arcmin square region
for flux density limits of 1~$\mu$Jy (upper) and 100~nJy (lower). The difference between the pair of images at the same flux density illustrates cosmic variance, with nearby large-lobed sources dominating the right-hand plots. There are 5 populations of sources shown: FRI galaxies (light, double-lobed), FRII galaxies (dark, double-lobed), starforming galaxies (dark, single disk), beamed FRIIs (dark, star), and FRIs (light, star).}
\label{fit1400sky}
\end{center}
\end{figure*}

\section{SKA SPECIFICATION FOR AGN CONTINUUM SURVEY SCIENCE}

Based on the SKA sky predictions presented here, if the source count
continues unchanged down to 100~nJy, there will be about 40 radio
sources per square arcmin. We have modelled these sources with finite
extents, such that these faint radio sources have dimensions
comparable with those of the optical discs of galaxies. At the full
sensitivity of the SKA, individual sources will appear blended at the
faintest flux density levels.

Resolutions of at least 0.05 arcsec at 1.4~GHz (i.e. 1000 km
baselines) are necessary to separate individual sources with a baseline distribution as specified in the SKA Science Requirements specification \cite{jo1}. However, to
image sources with sufficient detail to determine if the emission comes from the
disk, star formation or an AGN will require higher resolutions.  It is
also critical that the SKA resolution at least matches that of
complementary facilities (e.g. JWST) for multi-wavelength
studies. Also, detailed follow-up of SKA continuum radio surveys will
be necessary for `high precision' cosmology: Planck and its
successor missions will require detailed foreground excision, where
radio-detected AGN are contaminants to the CMB signal.

Finally, it is vital that the SKA is capable of making simultaneous
wide-band observations. These will be used to monitor the
time-variability of radio sources, furthering the study of the
mechanisms driving intrinsic as well as external variability, e.g. the
intra-day variability (IDV) phenomenon.

\end{document}